\documentclass[10pt,conference]{IEEEtran}
\usepackage{graphicx}
\usepackage{epsfig}
\usepackage{latexsym}
\usepackage{amsfonts}
\usepackage{here}
\usepackage{rawfonts}
\usepackage[latin1]{inputenc}
\usepackage[T1]{fontenc}
\usepackage{calc}
\usepackage{capitalgreekitalic}
\usepackage{url}
\usepackage{enumerate}
\usepackage{color}
\usepackage[tbtags]{amsmath}
\usepackage{amssymb}
\usepackage{upref}
\usepackage{epic,eepic}
\usepackage{times}
\usepackage{dsfont}
\usepackage{comment}
\usepackage{cite}

\newcommand{\qed}{\nobreak \ifvmode \relax \else
  \ifdim\lastskip<1.5em \hskip-\lastskip
  \hskip1.5em plus0em minus0.5em \fi \nobreak
  \vrule height0.75em width0.5em depth0.25em\fi}

\newcounter{step}
\newlength{\totlinewidth}
  {\end{list}%
  \rule{\linewidth}{1pt}}
\newcounter{substep}

  {\end{list}}

\newlength{\aligntop}
\newlength{\alignbot}
 \makeatletter
\renewenvironment{align}{%
  \vspace{\aligntop}
  \start@align\@ne\st@rredfalse\m@ne
}{%
  \math@cr \black@\totwidth@
  \egroup
  \ifingather@
    \restorealignstate@
    \egroup
    \nonumber
    \ifnum0=`{\fi\iffalse}\fi
  \else
    $$%
  \fi
  \ignorespacesafterend%
  \vspace{\alignbot}\par\noindent
} \makeatother

\IEEEoverridecommandlockouts

\begin{document}
\title{On the Effect of I/Q Imbalance on Energy Detection and a Novel Four-Level Hypothesis Spectrum Sensing\vspace*{-0em}}
\author{
\authorblockN{Omid Semiari, \emph{Student Member, IEEE}, Behrouz Maham, \emph{Member, IEEE}, and Chau Yuen, \emph{Senior Member, IEEE}}
    \thanks{Copyright (c) 2013 IEEE. Personal use of this material is permitted. However, permission to use this material for any other purposes must be obtained from the IEEE by sending a request to pubs-permissions@ieee.org.
    Omid Semiari and Behrouz Maham are with Electrical and Computer Engineering department, University of Tehran, Iran, e-mails: \protect\url{{omid.semiari, bmaham}@ut.ac.ir}. Chau Yuen is with Singapore University of Technology and Design, Singapore, e-mail: \protect\url{yuenchau@sutd.edu.sg}. This work was supported in part by the Singapore University Technology and Design (grant no. SUTD-ZJU/RES/02/2011), and in part by the Iran Telecommunications Research Center (ITRC).}}
\maketitle
\vspace*{-.6em}
\begin{abstract}
Direct-conversion transceivers are in demand, due to low
implementation cost of analog front-ends. However, these
transmitters or receivers introduce imperfections such as in-phase
and quadrature-phase (I/Q) imbalances. In this paper, we first
investigate the effect of I/Q imbalance on the performance of
primary system, and show that these impairments can severely degrade
the performance of cognitive radio system that are based on
orthogonal frequency division multiplexing (OFDM) multiple access
scheme. Next, we design a new four-level hypothesis blind detector
for spectrum sensing in such cognitive radio system, and show that
the proposed detector is less vulnerable to I/Q imbalance than
conventional two-level detectors.
\end{abstract}
\vspace*{-0em}
\section{Introduction}
Cognitive radio is a promising approach to alleviate the spectrum
scarcity faced by today wireless communication systems. The spectrum
utilization has been noticed to be less than $10\%$ in practice
\cite{o1}, and thus leads to the idea of reusing spectrum by a
secondary network\cite{oo2}.
In a cognitive radio network, secondary or unlicensed users benefit
by opportunistically accessing the spectrum, while primary or
licensed users are protected from the interference.

The recent development in integrated circuit (IC) technology and
incorporation of complete phase-locked loop devices in low-cost IC
packages have made direct-conversion transceivers widely accepted.
Nevertheless, in contrary to their desirable characteristics, the implementation of
direct-conversion transceivers causes various impairments associated
with the analog components. The in-phase and quadrature-phase (I/Q)
imbalance has been known as a major source of analog impairments in
high-speed wireless communication systems\cite{oo20}.

The effect of I/Q imbalance on traditional communication systems
have been comprehensively investigated in previous works
\cite{oo223,review1,review2,review3,review4,review5,review6}, and
various compensation algorithms have been proposed \cite{oo5, oo25,
oo26, oo27}. However, many of the proposed algorithms cannot be
applied to cognitive radio systems. For instance, a compensation
method for I/Q distortions based on a novel pilot pattern is
proposed in\cite{oo27}. Since there is no cooperation between the
primary transmitter and the secondary receiver in a cognitive radio
network, this method and other algorithms evolved out of channel
estimation techniques are not applicable in a cognitive radio
system. Although the negative effects of I/Q imbalance can be
severe, there has not been sufficient studies on the impact of I/Q
imbalance in cognitive radio system, and the way to mitigate it. In \cite{last1,last3}, Neyman-Pearson detector \cite{last2} have been exploited to perform spectrum sensing in the presence of transceiver  I/Q imbalance. Our work differs as we propose a multi-level detector and study the problem through minimum average cost detection \cite{poor} criterion.

In this paper, we consider the effect of interference imposed by I/Q
imbalance on the performance of the primary system with blind
spectrum sensing at the secondary user. By blind spectrum sensing,
we mean that the secondary user does not have access to the
instantaneous channel state information of the primary link, which
is the same assumption as in \cite{oo6}.
In order to tackle the effects of I/Q imbalance, we
propose a four-level hypothesis blind detector \textit{for the
first time}, and compare the performance of the proposed detector to the conventional two-level detector under the scenario of I/Q
imbalance impaired transceivers.

The subsequent sections of this paper are organized as follows.
Section II describes the system model under the effect of I/Q
imbalance and explains the idea of utilizing periodogram detector.
In Section III, we analyze the effects of
secondary transmitter I/Q imbalance on outage probability of the primary system. In Section IV, we explain how I/Q imbalance can affect the spectrum
sensing by the secondary user.
A four-level hypothesis blind detector is proposed to address the
problems of I/Q imbalance, and the detection problem is formulated.
Section V compares the performance of the new detector with conventional two-level detectors. Conclusion is given in Section VI.
\vspace*{-0em}
\section{System Model}
\label{sec:1}
We consider a cognitive radio network where the primary system
utilizes OFDMA technique for its uplink transmission. We assume an
overlay cognitive radio system in which the secondary user should
not interfere with primary user channels and access only vacant
channel. The primary system consists of $ U $ primary users and
dedicates total number of $ K $ subcarriers to them. Each primary
user transmits a set of designated data symbols $ s_{u,k} $ for
$k \in  \{-K/2, -K/2+1, ..., K/2\}  $\,and $ u\in \lbrace {1, 2,
..., U} \rbrace $ \!, where $ s_{u,k} $ is taken from an\!\!\!
\,\emph{ M}-PSK symbol constellation and only one user transmit on a
single subcarrier, i.e., orthogonal multiuser transmission. Without
loss of generality, we discard the user index \emph{u}, since we
assume the same modulation scheme for all primary users, i.e.,
\emph{ M}-PSK symbol constellation. The average power of each
subcarrier is assumed to be $P_{k}$, $k \in  \{-K/2, -K/2+1, ...,
K/2\}  $\,. We assume that no data is transmitted on the direct
current (DC) subcarrier and \emph{K} is an even number. The cyclic
prefix is assumed to be longer than the impulse response of the
channel and hence, there is no inter-carrier interference. In addition,
we assume that the secondary user knows \emph{a priori} information
about potential primary systems including frame structure,
subcarrier structure like the Discrete Fourier Transform (DFT) size,
the transmission parameters (fundamental symbol rate, cyclic prefix
length), and the level of interference tolerance of primary systems
\cite{oo8}. After removing the cyclic prefix and DFT operation, the
channel in the subcarrier \emph{k} can be represented as a complex
channel $ h_k $, which is modeled as Rayleigh fading with
independent Gaussian distributed random variable of variance $
\sigma_{ h_k }^2 $. It has been shown that I/Q imbalance in
OFDM-based systems results in a mutual interference between
subcarriers that are located symmetrically to the DC carrier
\cite{oo9,oo10}. In this paper, we focus on transmitter's I/Q
imbalance, since the direct-conversion architecture is commonly
employed in transmitters\cite{oo11}. We assume that both primary and
secondary direct-conversion transmitters are wideband, in which the
inter user interference imposed by I/Q imbalance is probable.

In practice, there is neither ideal amplitude matching nor exact $
\pi/2 $ phase difference among I and Q signals.  Thus, the actual
transmitted signal $ x_k $, which is different from the ideal
transmitted signal, can be modeled as \cite{oo28}:\vspace*{-0em}
\begin{equation}\label{eq:1}
    {x}_{k}=\alpha_{T}\sqrt{P_{k}}s_{k} + \beta_{T}\sqrt{P_{-k}}s_{-k}^{\dag},
\end{equation}\vspace*{-0em}
where $ {( .)}^{\dag}$ denotes the complex conjugate. Moreover, the
signals sent by primary users are assumed to be normalized, i.e.,
$\mathrm{E}\{|s_{k}|\}=\mathrm{E}\{|s_{-k}|\}=1$. From (\ref{eq:1}), at the
primary transmitter, there is an interference induced by I/Q
imbalance, which can be a significant performance limiting factor,
especially in the mobile uplinks. The degree of I/Q imbalance can be
evaluated in terms of an image rejection ratio (IRR), which is
defined by ${A}_{T}=\frac{{\vert \beta_{T} \vert}^2 }{{\vert
\alpha_{T} \vert}^2 }$ and ${A}_{R}=\frac{{\vert \beta_{R} \vert}^2
}{{\vert \alpha_{R} \vert}^2 }$ for transmitter and receiver,
respectively \cite{oo29}.
Parameters $ \alpha_T$ and $ \beta_T$ in (\ref{eq:1}) are two
complex scalars given by\vspace*{-0em}
\begin{align}\label{eq:2}
   \alpha_{T}=\cos(\theta_{T})+j\epsilon_{T}\sin(\theta_{T}), \,\, \beta_{T}=\epsilon_{T}\cos(\theta_{T})-j\sin(\theta_{T}),
\end{align}
and $ \epsilon_T$ and $ \theta_T $ represent the amplitude and phase
mismatch of transmitter I/Q signals, respectively \cite{oo12}. In addition, $ \alpha_R$ and $ \beta_R$ are defined based on (\ref{eq:2}), but with respect to the receiver mismatch parameters. From now, we use subscripts ${T,p}$ and $T,s$ to differentiate between the transmit I/Q imbalance parameters of the primary user and that of the secondary user, respectively. From
(\ref{eq:1}), the received signal at secondary receiver can be written as
\begin{equation}\label{eq:3}\vspace*{-0em}
   y_{k}=\alpha_{T,p}h_{k}\sqrt{P_{k}}s_{k}+\beta_{T,p}h_{k}\sqrt{P_{-k}}s_{-k}^{\dag}+\omega_{k},
\end{equation}
where $ \omega_k $ is the noise with independent and identically distributed (i.i.d.) Gaussian distribution with zero mean and variance $\sigma_n^2$, i.e., $ \omega_k \sim {N(0,\sigma_n^2)}$.


A typical energy detector consists of a low pass filter to remove
the out-of-band noise and adjacent interference, an analog to
digital converter, as well as a square law device to compute the
energy. However, this implementation computes the energy of signal
within the entire frequency band that is not a preferable approach
to perform the spectrum sensing\cite{oo14}. Hence, a periodogram
solution is proposed in\cite{oo14}, which is depicted in Fig.1.
In this figure, indices $i$ and
$k$ represent the time domain and frequency domain, respectively.
\begin{figure}
  \centering
  \includegraphics[width=\columnwidth]{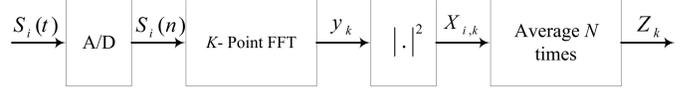}\\
  \caption{Block diagram of a priodogram detector for OFDM-based spectrum sensing.\vspace{-2em}
  }\label{f0}
\end{figure}
From (\ref{eq:3}), the output of periodogram detector is given by
\begin{align}\label{eq:4}
    Z_{k}=\frac{1}{N}\sum_{i=1}^{N}{|y_{k,i}|^2}=\frac{1}{N}\sum_{i=1}^{N}{X_{k,i}}=\frac{1}{N}\sum_{i=1}^{N}{(y_{R,i}^2+y_{I,i}^2)}
\end{align}
where $y_{R}$ and $y_{I}$ denote real and imaginary parts of
$y_{k}$, respectively. In addition, variables $N$ and $X_{k,i}$
represent the number of time-domain data packets and the square-law detector's output for the \emph{i}-th data packet
of \emph{k}-th subcarrier, respectively.\vspace*{-0em}
\section{I/Q Imbalance Effect on Primary System Performance Metric}
\label{sec:3-2}
In this section, we analyze the effect of interference imposed by
secondary user on the performance of primary system. This
interference occurs due to the I/Q imbalance of the secondary user
transmitter. Fig.\,2 demonstrates a cognitive radio network with an
OFDMA uplink transmission of primary users. Meanwhile, the
secondary user senses subcarrier \emph{k} in order to decide if it
is vacant. In this example, neither primary user 1 nor primary user
2 transmits data on subcarrier \emph{k}, while the primary user 2
transmits over subcarrier \emph{-k}. The two-level energy detector's decision for subcarrier $k$ is based only on the signal's averaged power in $k$-th subcarrier. In fact, although the secondary user may be aware of the transmit I/Q imbalance, it just incorporates the information derived from the $k$-th output of the FFT block for deciding on subcarrier $k$, regardless of the state of subcarrier $-k$. Hence, if the secondary user
uses two-level detector for spectrum sensing, it would transmit over
subcarrier \emph{k}, then as a result, primary user 2 would be impaired by the
I/Q imbalance from secondary user. As shown in Fig.\,2, the
secondary transmission on an idle subcarrier, e.g. subcarrier $k$,
introduces an interference on subcarrier $-k$. The primary receiver
treats the secondary interference as noise. Hence, the received
signal at the primary receiver is
\begin{align}\label{eq:5}
 y_{p,-k}=\sqrt{P_{-k}}s_{-k}g_{-k}+\beta_{T,s}\sqrt{P_{0}}s_{s,k}^{\dag}h_{-k}+w_{p}
\end{align}
where $g_{-k}$ denotes the channel coefficient of the primary link
on subcarrier $-k$, $P_0$ is the transmitted power of the secondary
user, and $s_{s,k}$ is secondary signal on subcarrier $k$. Moreover,
$w_p$ is the additive Gaussian noise of the primary receiver with
zero mean and variance of $N_p$, i.e., $ w_{p} \sim {N(0,N_{p})}$
and is independent of channel coefficients.
\begin{figure}
  \centering
  \includegraphics[width=\columnwidth]{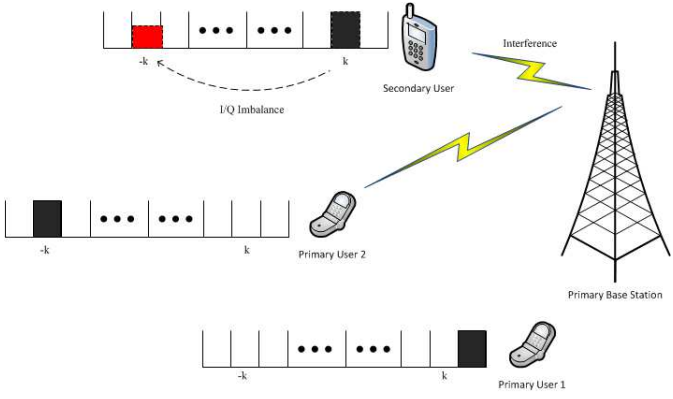}\\
  \caption{An example of a secondary user, interfering with OFDMA primary users uplink transmission due to I/Q imbalance caused by cognitive radio system transceivers.\vspace{-2em}
  }\label{f0}
\end{figure}
Assuming a primary uplink transmission with a fixed transmission rate of $R_{p}$, the outage probability is defined as $\rho_{0}\triangleq \{r_{p}<R_{p} \}$ \cite{oo16}. From (\ref{eq:5}), and by considering the normalized bandwidth, the data rate $r_{p}$ is given by
\begin{align}\label{eq:6}
 r_{p}=\log_{2}(1+\gamma)=\log_{2}\left(1+\frac{P_{-k}|g_{-k}|^2}{N_{p}+|\beta_{T,s}|^{2}P_{0}|h_{-k}|^2}\right)\,,
\end{align}
where $\gamma$ denotes the signal to interference plus noise ratio (SINR). The outage probability can then be expressed as
$\rho_{0}=~\text{Pr}\{\gamma<\gamma_{th} \}$, where\vspace{0cm}
$\gamma_{th}=2^{R_p}-1$. Hence,
\begin{align}\label{eq:7}
 \rho_{0}&=\text{Pr}\left\{\frac{X_1}{1+X_2}<\gamma_{th}\right\}\notag\\
 &=\int_{-\infty}^{\infty}\text{Pr}\{X_1<\gamma_{th}(1+X_2)|X_2\}f_{X_2}(x_2)dx_2\, ,
\end{align}
where $X_1$ and $X_2$ are two exponential random variables with mean $\sigma_{X_1}^2=\frac{P_{-k}}{N_p}\sigma_{g_{-k}}^2$ and $\sigma_{X_2}^2=\frac{|\beta_{T,s}|^{2}P_{0}}{N_p}\sigma_{h_{-k}}^2$, respectively. Therefore, the outage probability at the primary receiver can be calculated as\vspace*{0em}
\begin{align}\label{eq:8}
 \rho_{0}&=\int_{0}^{\infty} \left(1-e^{-\frac{\gamma_{th}(1+x_2)}{\sigma_{X_1}^2}}\right)\left(\frac{1}{\sigma_{X_2}^2}e^{-\frac{x_2}{\sigma_{X_2}^2}}\right)dx_2\notag\\
 &=1-\left(\frac{\sigma_{X_1}^2}{\sigma_{X_1}^2+\sigma_{X_2}^2}\right)e^{-\frac{N_p}{\sigma_{X_1}^2}\gamma_{th}}\,\, .
\end{align}

\vspace*{-0em}
\section{Four-Level Hypothesis Test for I/Q Imbalance Impaired Spectrum Sensing}
In this section, we introduce a four-level hypothesis test, and
apply Bayesian minimum average cost criterion to this model\cite{poor}. To
understand the motivation of the proposed four-level hypothesis
test, again consider the scenario depicted in Fig.\,2.

In addition to the problem discussed in previous section, I/Q
imbalances imposed by primary users can make a secondary user assume
a vacant subcarrier busy. This would rise the probability of false
alarm and degrades the system performance. Therefore, we use this
notion to propose a four-level hypothesis test, where the detector
can adjust the decision thresholds more intelligently and detect
vacant subcarriers more precisely. We model the four possible cases that could happen at a secondary receiver as follows:
  \begin{equation*}\vspace*{-0em}
  \begin{cases}
  \text{H}_{0}\text{:\,Only Noise}\\\vspace*{-0.3em}
  \text{H}_{1}\text{:\,I/Q Imbalance+Noise}\\\vspace*{-0.3em}
  \text{H}_{2}\text{:\,Primary user signal+Noise}\\\vspace*{-0.3em}
  \text{H}_{3}\text{:\,Primary user signal+I/Q Imbalance+Noise}.\vspace*{.8em}
  \end{cases}
  \end{equation*}
The first state occurs when the secondary user receives only noise
and there is no data transmission by any of the primary users in
both \emph{k}-th and \emph{-k}-th subcarriers. The second state
happens when there is transmission in subcarrier \emph{-k} by one of
the primary users, while subcarrier \emph{k} is left vacant. Hence,
the secondary user senses the interference plus noise, where the
interference is due to the I/Q imbalance of secondary receiver
or primary transmitter or both of them. If primary user
transmits in subcarrier \emph{k}, then secondary user senses only
primary signal plus noise as stated by the third hypothesis. The
forth state stands for the transmission of data over both
\emph{k}-th and \emph{-k}-th subcarriers by primary users.
Therefore, at this state the secondary user receives a signal
associated with the \emph{k}-th subcarrier and also an interference
plus noise.

The proposed detector is
able to differentiate the interference imposed by mirrored subcarrier from
the main signal, i.e., by detecting $\text{H}_{1}$. This is important, because secondary user can use this group of subcarriers in case that secondary transmitter does not suffer from considerable
I/Q imbalance. Even for the secondary user with high transmit I/Q imbalance, this detector helps user to be aware of the primary transmission on subcarrier $-k$ and avoid interfering with primary system.
According to the four-level hypothesis model and the expression for
the received signal from (\ref{eq:3}) and (\ref{eq:4}), the hypothesis test can be
formulated as\vspace*{0.3em}
   \begin{align}\label{eq:9}
  \begin{cases}
  \text{H}_{0}:Z_{k}=&\!\!\!\!\frac{1}{N}\sum_{i=1}^{N}{|\omega_{k,i}|^{2}}\\
  \text{H}_{1}:Z_{k}=&\!\!\!\!\frac{1}{N}\sum_{i=1}^{N}{|(\beta_{T,p}h_{k,i})\sqrt{P_{-k,i}}s_{-k,i}^{\dag}+\omega_{k,i}|^{2}}\\
  \text{H}_{2}:Z_{k}=&\!\!\!\!\frac{1}{N}\sum_{i=1}^{N}{|(\alpha_{T,p}h_{k,i}\!)\sqrt{P_{k,i}}s_{k,i}+\omega_{k,i}|^{2}}\\
  \text{H}_{3}:Z_{k}=&\!\!\!\!\frac{1}{N}\sum_{i=1}^{N}|(\alpha_{T,p}h_{k,i}\!)\sqrt{P_{k,i}}s_{k,i}\\
   &\!\!\!\!+(\beta_{T,p}h_{k,i})\sqrt{P_{-k,i}}s_{-k,i}^{\dag}+\omega_{k,i}|^{2}.
  \end{cases}
  \end{align}\vspace*{0em}

From (\ref{eq:3}), it is obvious that the received signal $y_{k}$ is a complex random variable. Hence, $y_{k}=y_{R,k}+jy_{I,k}$ where\vspace*{-0em}
\begin{align}\label{eq:10}
y_{R,k}&=Re\{\alpha_{T,p}\sqrt{P_{k}}s_{k}+\beta_{T,p}\sqrt{P_{-k}}s_{-k}^{\dag} \}h_{R,k}\notag\\
&-Im\{\alpha_{T,p}\sqrt{P_{k}}s_{k}+\beta_{T,p}\sqrt{P_{-k}}s_{-k}^{\dag} \}h_{I,k}+\omega_{R,k},\notag\\
y_{I,k}&=Re\{\alpha_{T,p}\sqrt{P_{k}}s_{k}+\beta_{T,p}\sqrt{P_{-k}}s_{-k}^{\dag} \}h_{I,k}\notag\\
&+Im\{\alpha_{T,p}\sqrt{P_{k}}s_{k}+\beta_{T,p}\sqrt{P_{-k}}s_{-k}^{\dag} \}h_{R,k}+\omega_{I,k},
  \end{align}
and $\omega_{R,k}$ and $\omega_{I,k}$ represent real and imaginary
parts of $\omega_{k}$, respectively. It can be checked from (\ref{eq:10})
that real and imaginary parts of $y_{k}$ are Gaussian random
variables with zero mean and identical variances. In addition, $y_{R,k}$ and $y_{I,k}$ are uncorrelated \cite{review10}.
Hence, they are independent and $y_{k}$ can be represented as a
complex Gaussian random variable. Thus, the probability density function
of their square is given by\cite{review8}
\begin{align}\label{eq:11}
f_{y_{R}^2}(x) = f_{y_{I}^2}(x)=
\begin{cases}
\frac{1}{\sqrt{2\pi x\sigma^2}}e^{-\frac{x}{2\sigma^2}}&\mbox{if $x \geq 0$},\\
 0&\mbox{if $x < 0$},
 \end{cases}
\end{align}
where $f_{X}(.)$ and $\sigma^2$ denote the probability density function of variable $X$ and the variance of either real or imaginary part of $y_{k}$, respectively. In addition, considering the fact that $y_{R}^2$ and $y_{I}^2$ are independent, and from (\ref{eq:11}), the probability density function of $X_{k}$ is
\begin{align}\label{eq:12}
  f_{X_{k}}(x)&=\!\!\int_{-\infty}^{\infty}\!\!\! f_{y_{R}^2}(x-s)\,f_{y_{I}^2}(s)\,ds\!\!=\!\!\int_{0}^{x}\!\!\frac{1}{2\pi \sigma^{2}}\frac{e^{-\frac{x}{2\sigma^{2}}}}{\sqrt{s(x-s)}}ds\notag\\
  &=\frac{1}{2\pi \sigma^{2}}e^{-\frac{x}{2\sigma^{2}}}\int_{0}^{1}\frac{2}{\sqrt{1-u^{2}}}du=\frac{1}{2 \sigma^{2}}e^{-\frac{x}{2\sigma^{2}}} .
  \end{align}
From (\ref{eq:12}), the conditional probability density function of $X_{k}$ can be represented as exponential distribution as follows:
\begin{equation}\label{eq:13}
f_{X}(x|H_{i}) =
\begin{cases}
\frac{1}{2\sigma_{i}^2}e^{-\frac{x}{2\sigma_{i}^2}} &\mbox{if $x \geq 0$},\\
0&\mbox{if $x < 0$},
\end{cases}
\end{equation}
where $ i=0,1,2,3 $ and
\begin{equation}\label{eq:14}
  \begin{cases}
\sigma_{0}^{2}=&\!\!\!\!\frac{1}{2}\sigma_{n}^{2}\\
\sigma_{1}^{2}=&\!\!\!\!\frac{1}{2}(|\beta_{T,p}|^{2}P_{-k}|s_{-k}|^{2}\sigma_{h_{k}}^{2})+\sigma_{0}^{2}\\
\sigma_{2}^{2}=&\!\!\!\!\frac{1}{2}(|\alpha_{T,p}|^{2}P_{k}|s_{k}|^{2})\sigma_{h_{k}}^{2}+\sigma_{0}^{2}\\
\sigma_{3}^{2}=&\!\!\!\!\frac{1}{2}(|\alpha_{T,p}|^{2}P_{k}|s_{k}|^{2}\\
&\!\!\!\!+2Re(\alpha_{T,p}\beta_{T,p}\sqrt{P_{k}}\sqrt{P_{-k}}s_{k}s_{-k}^{\dag}))\sigma_{h_{k}}^{2}+\sigma_{1}^{2}\,.
  \end{cases}
  \end{equation}
The periodgram detector, introduced in Section II, computes the average of FFT output for \emph{N} time domain windowed data packets for each frequency slot. The overall output of the detector for each frequency slot is a Gamma distributed random variable, since it is a sum of \emph{N} exponential random variables with identical means, i.e., $Z_{k}=\frac{1}{N}\sum_{i=1}^{N}|y_{i,k}|^2$. Therefore, we have
\begin{align}\label{eq:15}
f_{Z_{k}}(z|H_{i})=
\begin{cases}
\frac{1}{\Gamma(N)}(\frac{1}{N\sigma_{i}^2})^{N}z^{(N-1)}e^{-\frac{z}{N\sigma_{i}^2}}\,\,\,&\mbox{if $z \geq 0$},\\
0\,\,\,&\mbox{if $z < 0$},
\end{cases}
\end{align}
where $\Gamma(.)$ is the gamma function\cite[Sec. (8.31)]{oo15}.

Now, we apply the minimum average cost criterion to this model. Without losing any generality, a uniform cost test with equal prior probabilities is assumed for all hypotheses. Thus, $ \text{H}_i $ is selected if\cite{poor}
\begin{equation}\label{eq:16}
   \frac{f_{Z_{k}}(z|\text{H}_{i})}{f_{Z_{k}}(z|\text{H}_{j})}>1  \text{ for } i,j = 0, 1, 2, 3 \text{ and } i\not= j\,.
\end{equation}
It can be shown that by applying the conditions in (\ref{eq:16}), four
decision regions are obtained as follows:
\begin{equation}\label{eq:17}
  \begin{cases}
\Lambda_{0}:\lbrace 0<z<S_{01}, 0<z<S_{02}, 0<z<S_{03}\rbrace\\
\Lambda_{1}:\lbrace S_{01}<z<S_{12}, S_{01}<z<S_{13}\rbrace\\
\Lambda_{2}:\lbrace S_{12}<z<S_{23}, S_{02}<z<S_{23}\rbrace\\
\Lambda_{3}:\lbrace S_{03}<z, S_{13}<z, S_{23}<z  \rbrace \,\,\,\,\,\, ,
  \end{cases}
  \end{equation}
where $ \Lambda_i $ is the corresponding region of each hypothesis and
\begin{equation}\label{eq:18}
   S_{ij}=S_{ji}=\frac{N^{2}\ln\frac{\sigma_{i}^2}{\sigma_{j}^2}}{\frac{1}{\sigma_{j}^2}-\frac{1}{\sigma_{i}^2}}  \text{ for } i,j = 0, 1, 2, 3 \text{ and } i\not= j .
\end{equation}
In Appendix A, it is shown that $ S_{01}<S_{02}<S_{03}$, $S_{02}<S_{12}<S_{13}$, and $S_{03}<S_{13}<S_{23}$. Thus, the expressions in (\ref{eq:17}) can be simplified as:
\begin{align}\label{eq:19}
&\Lambda_{0}:\lbrace 0<z<S_{01}\rbrace,
\,\,\,\,\,\,\,\,\Lambda_{1}:\lbrace S_{01}<z<S_{12}\rbrace,\notag\\
&\Lambda_{2}:\lbrace S_{12}<z<S_{23}\rbrace,
\,\,\Lambda_{3}:\lbrace 0<S_{23}<z\rbrace.
  \end{align}
In order to analyze the effect of I/Q imbalance, we calculate the
probability of false alarm, i.e., when there is no signal
transmission in subcarrier \emph{k} , but secondary user assumes
that the channel is busy. Equivalently,
\begin{align}\label{eq:20}
P_{fa}&=P(\text{H}_{2}|\text{H}_{1})+P(\text{H}_{2}|\text{H}_{0})+P(\text{H}_{3}|\text{H}_{1})+P(\text{H}_{3}|\text{H}_{0})\notag\\
&=\!\!\int_{\Lambda_{2}}\!\![P(z|\text{H}_{1})\!+\!P(z|\text{H}_{0})]\,dz\!\!+\!\!\!\int_{\Lambda_{3}}\!\![P(z|\text{H}_{1})\!\!+\!\!P(z|\text{H}_{0})]\,dz .
\end{align}
From (\ref{eq:15}), the integrals can be evaluated using the following:
\begin{align}\label{eq:21}
 P(\text{H}_{3}|\text{H}_{i})&\!=\!\!\int_{\Lambda_{3}}\!\!\!P(z|\text{H}_{i})\!\!=\!\!\int_{S_{23}}^{\infty}{\frac{1}{\Gamma(N)}(\frac{1}{N\sigma_{i}^2})^{N}\!z^{(N-1)}e^{-\frac{z}{N\sigma_{i}^2}}}dz\notag\\
 &=\frac{1}{\sigma_{i}^{2}\Gamma(N+1)}\int_{S_{23}}^{\infty}{(\frac{z}{N\sigma_{i}^2})^{N-1}}e^{-(\frac{z}{N\sigma_{i}^2})}dz\notag\\
 &=\frac{1}{\Gamma(N)}\int_{\frac{S_{23}}{N\sigma_{i}^2}}^{\infty}{t^{N-1}e^{-t}}dt=
 \frac{\Gamma(N, \frac{S_{23}}{N\sigma_{i}^2})}{\Gamma(N)},
\end{align}
where $\Gamma(.,.)$ is the incomplete gamma function\cite{oo15}. Therefore, from (\ref{eq:20}) and (\ref{eq:21}) the probability of false alarm can be expressed as
\begin{align}\label{eq:22}
 P_{fa}&\!\!=\!\!\frac{\Gamma(N, \frac{S_{12}}{N\sigma_{1}^2})-\Gamma(N, \frac{S_{23}}{N\sigma_{1}^2})}{\Gamma(N)}+\frac{\Gamma(N, \frac{S_{12}}{N\sigma_{0}^2})-\Gamma(N, \frac{S_{23}}{N\sigma_{0}^2})}{\Gamma(N)}\notag\\
 &\!+\!\frac{\Gamma(N, \frac{S_{23}}{N\sigma_{1}^2})}{\Gamma(N)}\!+\!\!\frac{\Gamma(N, \frac{S_{23}}{N\sigma_{0}^2})}{\Gamma(N)}\!=\!\frac{\Gamma(N, \frac{S_{12}}{N\sigma_{1}^2})}{\Gamma(N)}\!+\!\frac{\Gamma(N, \frac{S_{12}}{N\sigma_{0}^2})}{\Gamma(N)}.\vspace{0cm}
\end{align}
Similarly, the probability of detection is
given by
\begin{align}\label{eq:23}
 P_{D}&=P(\text{H}_{2}|\text{H}_{2})+P(\text{H}_{3}|\text{H}_{3})\notag\\
 &=\frac{\Gamma(N, \frac{S_{12}}{N\sigma_{2}^2})}{\Gamma(N)}-\frac{\Gamma(N, \frac{S_{23}}{N\sigma_{2}^2})}{\Gamma(N)}+\frac{\Gamma(N, \frac{S_{23}}{N\sigma_{3}^2})}{\Gamma(N)}\,\,.
\end{align}

It must be noted that different side information is needed in order
to calculate the detection thresholds. For instance, the statistics
of the interference link between primary users and secondary users,
i.e. $\sigma_{h_{k}}^2$, could be inferred by listening to the
downlink transmission of the primary system\cite{oo16}. In addition,
the determination of primary transmitter IRR values or transmit
power requires either explicit signaling from the primary system to
the secondary users, or the primary transmit power could be
estimated if secondary user roughly knows about its distance from
the nearby primary user. This work could also be extended to MIMO
systems in which localization and other array-antenna techniques are
incorporated.
\vspace*{-.8em}
\section{Simulation Results}
\begin{figure}
  \includegraphics[width=\columnwidth]{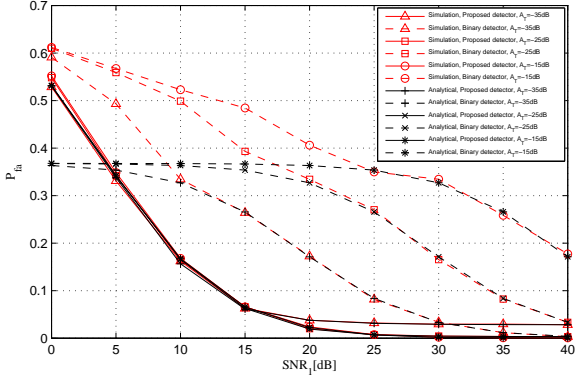}\ 
\caption{The comparison between the probability of false alarm of the proposed four-level detector and the conventional two-level detector for different values of IRRs. The dashed and solid curves represent the probability of false alarm for conventional detector and proposed detector, respectively. The
results are obtained for different values of IRRs and for the
constant transmitted power of subcarrier $-k$, i.e.,
$\text{SNR}_{2}=-10dB$.}\vspace{-1em}
\label{fig:3}       
\end{figure}

\begin{figure}
  \includegraphics[width=\columnwidth]{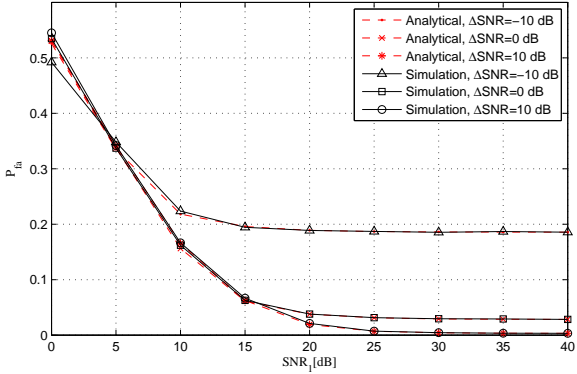}\ 
\caption{The probability of false alarm for different $\Delta \text{SNR}$ values, when IRR= -15dB. The dashed and solid curves represent the analytical
results and the simulations, respectively.  }\vspace{-1.5em}
\label{fig:4}       
\end{figure}
\label{sec:1} In this section, we provide simulations to evaluate
the analytical results provided in previous sections. We assume a
cognitive radio network with an OFDMA primary system with four
users, i.e., $U=4$. Moreover, the total number of subcarriers is
$K=512$, i.e., 128 subcarrier per user with 16-PSK constellation. We also assume a secondary
user that uses blind detection for spectrum sensing, based on the
four-level hypothesis test proposed in Section IV. The magnitude of
the frequency domain channel coefficients $h_{k}$ and $h_{-k}$ are
assumed independent Rayleigh distributed random variables with
variance $\sigma_{h_{k}}^{2}=\sigma_{h_{-k}}^{2}=1$.
In order to examine the statistical aspects of energy detector,
simulations are done over 20 million samples and for one sample
package, i.e., $N=1$. Here, we define the transmitted signal to
noise ratio for subcarrier $k$ and $-k$ as
$\text{SNR}_{1}=\frac{P_{k}|s_{k}|^{2}}{\sigma_{n}^2}$ and
$\text{SNR}_{2}=\frac{P_{-k}|s_{-k}|^{2}}{\sigma_{n}^2}$,
respectively. We consider orthogonal signaling, hence interference is only due to the I/Q imbalance. Secondary receiver treats primary interference as the Gaussian noise, since it is not completely aware of the primary signal's modulation parameters. The SNR at each subcarrier can be determined by priodogram detector described in Section II.



The comparison between the proposed four-level detector and the
conventional two-level detector is shown in Fig.3. Based on the results, the proposed detector is less vulnerability to the I/Q imbalance
effect than the two-level detector. Since the two-level detector
sets its detection threshold based only on the estimated noise
variance, the interference imposed by other subcarriers can fool the
detector. On the other hand, the proposed detector uses the joint
information of both subcarriers $k$ and $-k$, by adding hypotheses
$\text{H}_{1}$ and $\text{H}_{2}$ into its detection procedure. This helps the detector
to sense the spectrum based on the variance of both noise and
interference which leads to lower probability of false alarm.

Fig.4 demonstrates both analytical and simulation results of I/Q
imbalance effect on probability of false alarm for the proposed
detector by considering various $\Delta \text{SNR}=10
\log\frac{\text{SNR}_{1}}{\text{SNR}_{2}}$, which denotes the signal to noise ratio difference between the
subcarriers $k$ and $-k$. Fig.4 shows that even
for $\Delta \text{SNR}=-10\text{dB}$, the probability of false alarm is
significantly increased.

The comparison of joint transmitter-receiver I/Q imbalance effect on
the probability of detection with the case when I/Q imbalance exists
only in transmitters is shown in Fig. 5. The joint
transmitter-receiver I/Q imbalance effect can be investigated based
on the procedure introduced in this paper. Indeed, it can be proven
that the variance of received signal may differ from that of
presented in Section IV, nevertheless the detector output
distributions are identical in both scenarios. Hence, the
probability of false alarm and the probability of detection could be
found accordingly. As expected from analytical results, effect of
I/Q imbalance on the receiver's probability of detection is serious
when direct-conversion transceivers are used. However,
direct-conversion receivers offer significant power saving, hence
they might be as preferable as direct-conversion transmitters to be
utilized in wireless communication devices\cite{oo11}.

Eventually, Fig. 6 depicts the impact of I/Q imbalance of the secondary transmitter on the outage probability of primary receiver. It clearly demonstrates that the I/Q imbalance in secondary transmitter can dramatically affect the performance of the primary system.
\vspace*{-0em}
\section{Conclusion}
In this paper, the effects of I/Q imbalance on both primary and
secondary systems have been investigated. It was explained how I/Q
imbalance can make the secondary user interfere to an OFDMA primary
system if the secondary user employs binary detector. Therefore, we
design a new detector based on four-level hypothesis for blind
spectrum sensing in an overlay cognitive radio system. The detection
algorithm jointly considers the information of DFT blocks output for
a specific subcarrier and the interference imposed by its
symmetrical counterpart.  Simulation results show that the proposed
detector decreases the probability of false alarm, while avoids
interfering to primary system.
\vspace*{-0.4em}
\section*{APPENDIX A}
From (\ref{eq:14}) and assuming $\lambda_{i}=\frac{1}{\sigma_{i}^2}$ as well as $\sigma_2^2>\sigma_1^2$, it is obvious that $0<\lambda_{3}<\lambda_{2}<\lambda_{1}<\lambda_{0}$. Moreover, from (\ref{eq:18}), we have
\begin{align}\label{eq:26}
 S_{ij}&=S_{ji}=N^{2}\frac{\ln(\lambda_{i})-\ln(\lambda_{j})}{\lambda_{i}-\lambda_{j}} \notag\\
 &\triangleq N^{2}\frac{\Delta{F(x)}}{\Delta x}\Big|_{\lambda_{i}, \lambda_{j}} \,\,\,\text{ for } i,j = 0, 1, 2, 3 \text{ and } i\not= j,
\end{align}
where $F(x)=\ln(x)$. Therefore, (\ref{eq:26}) suggests that $S_{ij}$ is of the form of $F(x)$ differentiation, i.e., $\frac{1}{x}$, which is a decreasing function of $x$. Hence, $ S_{01}<S_{02}<S_{03}$, $S_{02}<S_{12}<S_{13}$, and $S_{03}<S_{13}<S_{23}$.
\label{sec:5-1}
\vspace{-.3cm}
\bibliographystyle{ieeetr}
\bibliography{references}

\begin{thebibliography}{10}

\bibitem{o1}
{Federal\,Communication\,Commissions}, {\em Facilitating Opportunities for
  Flexible, Efficient, and Reliable Spectrum use Employing Cognitive Radio
  Technologies}.
\newblock in FCC-03-322, 2003.

\bibitem{oo2}
F.~K. Jondral and T.~A. Weiss, ``Spectrum pooling: An innovative strategy for
  the enhancement of spectrum efficiency,'' {\em IEEE Radio Communications},
  vol.~42, no. 3, pp.~S8--S14, March 2004.

\bibitem{oo20}
C.~L. Liu, ``Impact of {I/Q} {I}mbalance on {QPSK-OFDM-QAM} {D}etection,'' {\em
  IEEE Trans. Consumer Electron.}, vol.~44, no.8, pp.~984--989, Aug. 1998.

\bibitem{oo223}
A.~Tarighat and A.~H. Sayed, ``{MIMO} {OFDM} receivers for systems with {I/Q}
  imbalances,'' {\em IEEE Trans. Signal Process.}, vol.~53, no.9,
  pp.~3583--3596, Sep. 2005.

\bibitem{review1}
M.~Krondorf and G.~Fettweis, ``Ofdm link performance analysis under various
  receiver impairments,'' {\em EURASIP Journal on Wireless Communications and
  Networking}, DOI:10.1155/2008/145279, 2008.

\bibitem{review2}
F.~Lopez-Martinez, E.~Martos-Naya, J.~F. Paris, and J.~Entrambasaguas, ``Exact
  closed-form {BER} analysis of {OFDM} systems in the presence of {IQ}
  imbalances and {ICSI},'' {\em IEEE Transactions on Wireless Communications},
  vol.~10, no.6, pp.~1914--1922, Jun. 2011.

\bibitem{review3}
Y.~Zou, M.~Valkama, N.~Ermolova, and O.~Tirkkonen, ``Analytical performance of
  {OFDM} radio link under {RX I/Q} imbalance and frequency-selective {R}ayleigh
  fading channel,'' in {\em Proc. IEEE Int. Workshop on Signal Processing Adv.
  Wireless Communications (SPAWC11)}, (San Francisco, CA), Jun. 2011.

\bibitem{review4}
S.~Mallick and S.~Majumder, ``Performance analysis of an {OFDM} system in the
  presence of carrier frequency offset, phase noise and timing jitter over
  {R}ayleigh fading channels,'' in {\em Proc. IEEE International Conf.
  Electrical Computer Engineering}, pp.~205--210, Dec. 2008.

\bibitem{review5}
R.~Zhang, E.~Au, and R.~Cheng, ``Impacts of {CFO}, {IQ} imbalance and phase
  noise on the system performance of {OFDM} systems,'' in {\em IEEE Int. Conf.
  on Communication Systems}, pp.~1041--1045, Nov. 2008.

\bibitem{review6}
S.~Krone and G.~Fettweis, ``Capacity analysis for {OFDM} systems with
  transceiver {I/Q} imbalance,'' in {\em Proc. IEEE GLOBECOM}, pp.~1--6, Nov.
  2008.

\bibitem{oo5}
Y.~Yoshida, K.~Hayashi, H.~Sakai, and W.~Bocquet, ``Analysis and compensation
  of transmitter {IQ} imbalances in {OFDMA} and {SC-FDMA} systems,'' {\em IEEE
  Trans. Signal Process.}, vol.~57, no. 8, pp.~3119--3129, Aug. 2009.

\bibitem{oo25}
M.~Valkama, A.~Shahed, L.~Anttila, and M.~Renfors, ``Advanced digital signal
  processing techniques for compensation of nonlinear distortion in wideband
  multicarrier radio receivers,'' {\em IEEE Trans. Microw. Theory Tech.},
  vol.~54, no. 6, pp.~2356--2366, Jun. 2006.

\bibitem{oo26}
J.~Tubbax, B.~Come, L.~V. der Perre, S.~Donnay, M.~Moonen, and H.~D. Man,
  ``Compensation of transmitter {I/Q} imbalance for {OFDM} systems,'' {\em In
  Proc. IEEE Int. Conf. Acoust., Speech, Signal Process. (ICASSP)},
  pp.~325--328, 2004.

\bibitem{oo27}
H.~A. Mahmoud, H.~Arslan, M.~K. Ozdemir, and F.~E. Retnasothie, ``{I/Q}
  imbalance correction for {OFDMA} uplink systems,'' {\em In Proc. IEEE Int.
  Conf. Commun. (ICC)}, Jun. 2009.

\bibitem{last1}
A.~ElSamadouny, A.~Gomaa, and N.~Al-Dhahir, ``Likelihood-based spectrum sensing
  of {OFDM} signals in the presence of {Tx/Rx} {I/Q} imbalance,'' {\em IEEE
  Global Communications Conference (GLOBECOM), Anaheim, CA}, vol.~57, no. 4,
  pp.~3616--3621, Dec. 2012.

\bibitem{last3}
A.~Gokceoglu, S.~Dikmese, M.~Valkama, and M.~Renfors, ``Enhanced energy
  detection for multi-band spectrum sensing under rf imperfections,'' {\em 8th
  International Conference on Cognitive Radio Oriented Wireless Networks
  (CROWNCOM), Washington, DC}, pp.~55--60, July 2013.

\bibitem{last2}
{S. M. Kay}, {\em Fundamentals of Statistical Signal Processing: Detection
  Theory.}
\newblock Princton, NJ, 1998.

\bibitem{poor}
{H. Vincent Poor}, {\em An Introduction to Signal Detection and Estimation}.
\newblock Princton, NJ, 1994.

\bibitem{oo6}
P.~De and Y.~C. Liang, ``Blind spectrum sensing algorithms for cognitive radio
  networks,'' {\em IEEE Trans. Veh. Technol.}, vol.~57, no. 5, pp.~2834--2842,
  Sep. 2008.

\bibitem{oo8}
K.~C. Cheng and R.~Prasad, {\em {C}ognitive {R}adio {N}etworks}.
\newblock John Wiley \& Sons, 2009.

\bibitem{oo9}
T.~C. Schenk, E.~Fledderus, and P.~F. Smudlers, ``{P}erformance {A}nalysis of
  {Z}ero-{I}{F} {MIMO} {OFDM} {T}ransceivers with {IQ} {I}mbalance,'' {\em
  Journal of Communications}, vol.~2, no. 7, pp.~9--19, Dec. 2007.

\bibitem{oo10}
Y.~Zou, M.~Valkama, and M.~Renfors, ``{A}nalysis and {C}ompensation of
  {T}ransmitter and {R}eceiver {I/Q} {I}mbalances in {S}pace-time {C}oded
  {M}ultiantenna {OFDM} {S}ystems,'' {\em EURASIP Journal on Wireless
  Communications and Networking}, doi:10.1155/2008/391025, 2008.

\bibitem{oo11}
A.~A. Abidi, ``Direct-conversion radio transceivers for digital
  communications,'' {\em IEEE Journal of Solid-State Circuits}, vol.~30, no.
  12, pp.~1399--1410, Dec. 1995.

\bibitem{oo28}
B.~Maham and O.~Tirkkonen, ``Transmit {A}ntenna {S}election {OFDM} {S}ystems
  with {T}ransceivers {I/Q} {I}mbalance,'' {\em IEEE Trans. Commun.}, vol.~60,
  no. 3, pp.~643--648, Mar. 2012.

\bibitem{oo29}
B.~Maham and A.~Hjorungnes, ``Impact of {T}ransceiver {I/Q} {I}mbalance on
  {T}ransmit {D}iversity of {B}eamforming {OFDM} {S}ystems,'' {\em IEEE Trans.
  Veh. Technol.}, vol.~61, no. 2, pp.~865--871, Feb. 2012.

\bibitem{oo12}
B.~Razavi, {\em {RF} {M}icroelectronic}.
\newblock Englewood Cliffs, NJ: Prentice-Hall, 1998.

\bibitem{oo14}
D.~Cabric, A.~Tkachenko, and R.~W. Brodersen, ``Spectrum {S}ensing
  {M}easurements of {P}ilot, {E}nergy, and {C}ollaborative {D}etection,'' {\em
  Proceeding of IEEE Military Communication Conference (MILCOM)}, Oct. 2006.

\bibitem{oo16}
B.~Maham, R.~Popovski, X.~Zhou, and A.~Hjoungnes, ``Cognitive {M}ultiple
  {A}ccess {N}etwork with {O}utage {M}argin in the {P}rimary {S}ystem,'' {\em
  IEEE Transactions on Wireless Communications}, vol.~10, no.10,
  pp.~3343--3353, Oct. 2011.

\bibitem{review10}
L.~Anttila, M.~Valkama, and M.~Renfors, ``Circularity-based {I/Q} imbalance
  compensation in wideband direct-conversion receivers,'' {\em IEEE
  TRANSACTIONS ON VEHICULAR TECHNOLOGY}, vol.~57, no. 4, pp.~2099--2113, July
  2008.

\bibitem{review8}
M.~K. Simon, {\em Probability Distributions Involving Gaussian Random
  Variables: A Handbook for Engineers, Scientists and Mathematicians}.
\newblock Springer-Verlag New York, Inc., Secaucus, NJ, 2006.

\bibitem{oo15}
I.~S. Gradshteyn and I.~M. Ryzhik, {\em Table of {I}ntegrals, {S}eries, and
  {P}roducts}.
\newblock San Diego, CA: Academic, 2000.

\end{thebibliography}

\begin{figure}
  \includegraphics[width=\columnwidth]{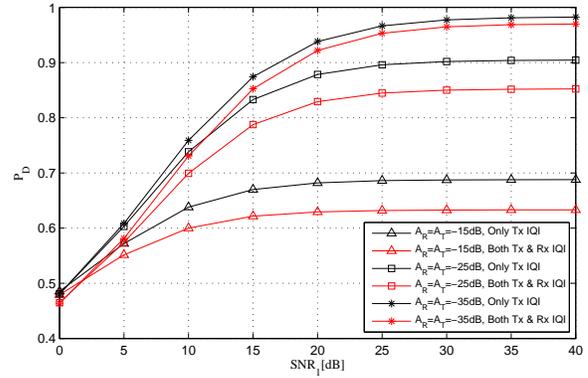}\ 
\caption{The comparison of joint transmitter-receiver I/Q imbalance effect on the probability of
detection with the case when I/Q imbalance exists only in
transmitters for different values of IRRs. The IRR values are assumed to be identical in transmitters and receivers, i.e., $A_T=A_R$.}\vspace{-1em}
\label{fig:7}       
\end{figure}

\begin{figure}
  \includegraphics[width=\columnwidth]{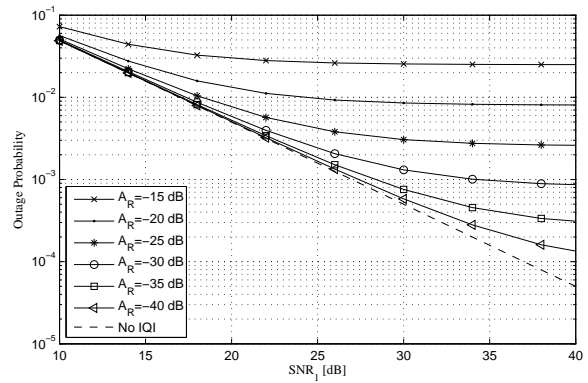}\ 
\caption{The effect of I/Q imbalance on outage probability of primary system for different values of IRRs}
\label{fig:1}       
\end{figure}
\end{document}